\documentclass[reprint, amsmath,amssymb,aps,pre,superscriptaddress,notitlepage]{revtex4-2}
\usepackage[normalem]{ulem}
\usepackage{xcolor}
\usepackage{graphicx}
\usepackage{bm}
\usepackage{url}
\usepackage{soul}
\usepackage{hyperref}
\usepackage{cancel}
\newcommand{\fsdel}[1]{{\color{teal}\ifmmode \xcancel{#1} \else \sout{#1} \fi}}

\newcommand{\ie}{\textit{i}.\textit{e}., }

\newcommand{\rem}[1]{}
\begin{document}
\title{Particle-Mediated Tuning of Defect Stability in Lamellar Block Copolymer Systems}
\author{Le Qiao}
\email{le.qiao@uni-mainz.de}
\affiliation{Institute of Physics, Johannes Gutenberg University Mainz, D55099 Mainz, Germany}
\author{Daniel A. Vega}
\affiliation{Instituto de Física del Sur (IFISUR), Consejo Nacional de Investigaciones Científicas y Técnicas (CONICET), Universidad Nacional del Sur, 8000 Bahía Blanca, Argentina}
\author{Friederike Schmid}
 \email{friederike.schmid@uni-mainz.de}
\affiliation{Institute of Physics, Johannes Gutenberg University Mainz, D55099 Mainz, Germany}
\date{\today}
\begin{abstract}
We study how colloidal inclusions modify the formation energy of dislocation pairs in lamellar block copolymer systems. Using a hybrid particle/Ginzburg--Landau model, we calculate defect formation energies by comparing defect-free and defect-containing states with and without embedded colloids. Finite-size scaling is used to obtain formation energies in the thermodynamic limit. The effect of colloid insertion depends strongly on particle sizes and surface patterning. Homogeneous particles increasingly stabilize dislocation pairs with increasing particle sizes. Particles larger than one lamellar domain preferentially occupy the dislocation cores, where they replace strained polymer rather than deforming defect-free lamellae. The magnitude of this stabilization depends on surface affinity. Balanced Janus particles instead increase the formation energy, because their competing surface preferences cannot be satisfied simultaneously near the curved core. Varying the patch ratio interpolates between these limits. Surface patterning has little effect for particles smaller than one lamellar domain but changes the formation energy by several tens of $k_BT$ for larger particles. These results provide quantitative guidelines for controlling topological defect stability in lamellar block copolymer systems.

\end{abstract}

\maketitle

\section{Introduction}
Block copolymers (BCPs) are versatile materials for self-assembled nanostructures, because microphase separation produces well-defined morphologies on nanometer length scales. Topological defects are intrinsic features of BCP ordering and can affect structural perfection, ordering pathways, and the mechanical, transport, and optical properties of the material.\cite{horvatSpecificFeaturesDefect2008,nagpalFreeEnergyDefects2012a,vegaOrderingMechanismsTwoDimensional2005b,liDefectsSelfAssemblyBlock2015,jangizehiDefectsDefectEngineering2020a} Although defects are often undesirable in applications requiring long-range periodic order, they can also be used as functional structural elements. Controlled defects can template nonperiodic nanopatterns for device fabrication.\cite{stoykovichDirectedAssemblyBlock2005a,stoykovichDirectedSelfAssemblyBlock2007,liuIntegrationDensityMultiplication2010} Defects in membrane and photonic materials can provide selective transport pathways or localized optical states.\cite{finkBlockCopolymersPhotonic1999} Defect cores may also exhibit thermodynamic behavior distinct from the surrounding ordered phase and induce local order--order transitions in thin copolymer films.\cite{abateDefectInducedOrderOrder2021}

Defect control in BCP films commonly relies on thermal or solvent annealing, which increases chain mobility and promotes equilibration. Recent shear--solvent annealing and in situ microscopy studies have demonstrated substantial defect reduction and revealed the strain-dependent dynamics of defect healing.\cite{kimShearsolvoDefectAnnihilation2019,murphyMappingDynamicsFluctuations2024} Annealing, however, primarily promotes global structural relaxation and offers limited control over the position, geometry, and persistence of individual defects. An alternative approach is to introduce dopants, including homopolymers, nanoparticles, grafted polymers, or small molecules, that modify the local free-energy landscape. This strategy is particularly relevant for lamellar morphologies. Bends, dislocations, and grain boundaries involve spatial variations in interfacial curvature and lamellar spacing, which generate local packing frustration and excess free energy.\cite{liDefectsSelfAssemblyBlock2015,hammondAdjustmentBlockCopolymer2003} 

In lamellar systems, dislocations play a central role in the mesoscale dynamics of defect evolution. This is because the motion and annihilation of orientational defects such as disclinations are mediated by the glide and climb of dislocations, making dislocations the primary carriers of topological rearrangements during coarsening.\cite{Harrison2000,Harrison2002} 
The two defect types also differ sharply in cost. An isolated disclination forces a compression of the lamellar spacing over an extended region, so its energy grows with the size of that region and it is either screened by arrays of dislocations or relaxed into curvature walls. An edge dislocation, by contrast, has a layer-displacement field and finite line tension independent of system size.\cite{klemanInteractionParallelEdge1974,pershanDislocationEffectsSmecticA1974} Dislocations are therefore the dominant and persistent sources of elastic distortion in lamellar systems, and understanding how additives modify their energetics is essential for predicting defect relaxation pathways. \cite{Harrison2002,Abate2016,Vu2018}

Dopants can preferentially segregate to these regions and modify their local packing, interfacial, and curvature costs. In the terminology of Li and Müller, such additives act as defectants: by analogy with surfactants that lower interfacial free energy through adsorption, defectants lower the excess free energy of localized nonbulk structures through selective segregation.\cite{liDefectsSelfAssemblyBlock2015} Dopants may therefore reduce the driving force for defect removal and alter defect kinetics by slowing the molecular rearrangements required for
annihilation.\cite{ginzburgKineticModelPhase1999a,hoheiselBlockCopolymernanoparticleHybrid2015a,ryuRetardationGrainGrowth2014}

Experiments provide evidence for this defectant mechanism. Burgaz and Gido found that T-junction grain boundaries, which are rare in neat lamellar BCPs because of their high excess energy, form abundantly in block-copolymer--homopolymer blends. The homopolymer enriches the strained junction region, accompanied by enlarged semicylindrical end caps and locally increased lamellar spacing.\cite{burgazTJunctionGrainBoundaries2000} Nanoparticles have likewise been shown to stabilize grain-boundary morphologies in lamellar BCP--nanoparticle blends.\cite{listakStabilizationGrainBoundary2006a,ryuRetardationGrainGrowth2014} Directed self-assembly studies further show that homopolymer partitions into sharply bent corners of nonregular patterns, where the neat copolymer is strongly frustrated.\cite{stoykovichDirectedAssemblyBlock2005a} Similarly, nanoparticles accumulate in the frustrated corners of guided bends, with spatial distributions reproduced quantitatively by self-consistent field theory.\cite{kangHierarchicalAssemblyNanoparticle2008} Together, these studies show that additives can localize in distorted and frustrated regions of BCP morphologies. 

A broad range of theoretical and computational approaches has been used to study dopant-mediated BCP self-assembly. Particle-resolved methods, including coarse-grained molecular dynamics,\cite{schultzComputerSimulationBlock2005,kalraEffectShearNanoparticle2010,kalraCoarsegrainedMolecularDynamics2009,burgos-marmolMolecularDynamicsJanus2021}
and Monte Carlo simulations,\cite{kangHierarchicalAssemblyNanoparticle2008,detcheverryMonteCarloSimulations2008,huhThermodynamicBehaviorParticle2000c,chervanyovConductivityDiblockCopolymer2022}
resolve local particle packing and polymer-mediated interactions, but their cost limits access to the mesoscopic scales required for isolated topological defects. Field-based approaches reach these scales more efficiently. Self-consistent field theory combined with particle density-functionals predicts equilibrium mesophases and particle distributions in diblock-copolymer--nanoparticle composites, showing that particle size and block selectivity control whether inclusions localize at domain centers or interfaces.\cite{matsenParticleDistributionsBlock2008,thompsonPredictingMesophasesCopolymerNanoparticle2001a,thompsonBlockCopolymerDirectedAssembly2002,sidesHybridParticleFieldSimulations2006} This size selectivity is also established experimentally: small enthalpically compatible particles segregate to block interfaces, whereas larger particles are driven toward domain centers by entropic effects.\cite{bockstallerSizeSelectiveOrganizationEnthalpic2003}

This framework has been extended to particles placed in idealized defect geometries held fixed in space, including X-, T-, and Y-shaped junctions. These calculations map particle-localization free-energy wells and show that their depth depends on defect symmetry and particle
size.\cite{kimFreeEnergyLandscape2014} Quantitative calculations of BCP defect free energies have also been performed with self-consistent field theory and single-chain-in-mean-field simulations, but mainly for neat copolymers in chemically guided or laterally confined geometries.\cite{liDefectsSelfAssemblyBlock2015,nagpalFreeEnergyDefects2012a,takahashiDefectivityLaterallyConfined2012a,hurMolecularPathwaysDefect2015a,liThermodynamicsKineticsDefect2016a,zhangHarnessingAnisotropicNanoposts2014} Hybrid particle--field approaches couple a continuum copolymer description to explicit colloids. The copolymer evolves by a time-dependent Ginzburg--Landau equation, often integrated with a cell-dynamics scheme,\cite{oonoStudyPhaseseparationDynamics1988,puriStudyPhaseseparationDynamics1988} and particle motion is described by Brownian dynamics.\cite{ginzburgModelingDynamicBehavior2000a,balazsMultiScaleModelBinary2000}
These methods have established how particle loading, size, shape, and chemical affinity affect equilibrium morphology and particle organization in BCP nanocomposites.\cite{diazCellDynamicSimulations2017,diazPhaseBehaviorBlock2018,diazCoassemblyJanusNanoparticles2019,diazHybridTimeDependentGinzburg2022,diazBlockCopolymerNanorod2020,diazNonsphericalNanoparticlesBlock2019a,diazLargeScaleThree2019} 

\begin{figure}[th]
    \centering
    \includegraphics[width=1\linewidth]{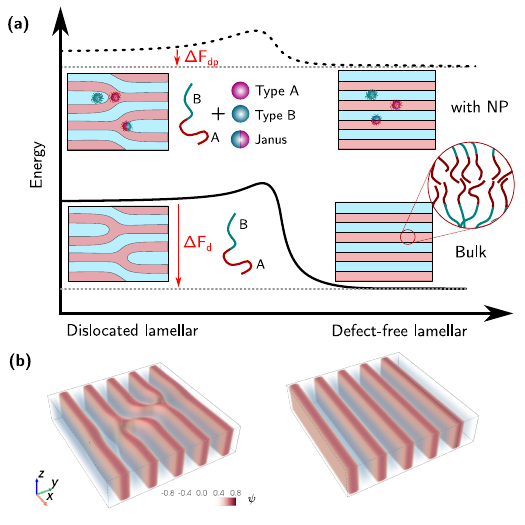}
\caption{Schematic of particle-mediated defect stabilization. (a)~Free-energy landscapes connecting the defect-containing lamellar state (left) and the defect-free lamellar state (right) for the undoped system (solid curve) and a system containing a particle pair (dashed curve). $\Delta F_d$ and $\Delta F_{dp}$ denote the corresponding dislocation-pair formation energies (Eqs.~\eqref{eq:dFd} and~\eqref{eq:dFdp}). As illustrated, particles that preferentially occupy the distorted core region can lower the formation energy relative to the undoped case. The three particle types considered here, A-selective, B-selective, and Janus particles, are shown schematically together with the diblock copolymer architecture. (b)~3D snapshots of order-parameter fields $\psi(\mathbf r)$ of the undoped system in the defect-containing (left) and defect-free (right) states, corresponding to the reference states in panel (a).}
\label{fig:sketch}
\end{figure}

In this work, we address a complementary question: We determine how colloidal dopants modify the formation energy of dislocation pairs in lamellar BCP systems (Fig.~\ref{fig:sketch}a), thereby potentially stabilizing them. This requires calculating the free-energy difference between defect-containing and defect-free states at equal particle content and an extrapolation to the thermodynamic limit: Lamellar dislocations produce long-ranged elastic distortions,\cite{ambrozicAnnihilationEdgeDislocations2004,klemanInteractionParallelEdge1974,pershanDislocationEffectsSmecticA1974} and opposite Burgers vectors interact over several lamellar
periods.\cite{liThermodynamicsKineticsDefect2016a} In a finite periodic cell, commensurability and periodic-image interactions therefore contribute to the apparent formation energy.

We use a hybrid Ohta--Kawasaki-type time-dependent Ginzburg--Landau (TDGL) model coupled to Brownian dynamics for explicit colloidal particles. The polymer field (Fig.~\ref{fig:sketch}b) is evolved with a semi-implicit Fourier-spectral scheme. We consider homogeneous particles with A- or B-selective surfaces and patchy particles with A- and B-selective surface regions. By comparing defect-free and defect-containing states at fixed particle content, we determine the particle-induced change in dislocation-pair formation energy. Finite-size scaling removes commensurability and periodic-image effects and yields thermodynamic-limit values. We thereby determine how particle size and surface patterning stabilize or destabilize dislocations.

\section{Models and Simulations}
\subsection{Time-dependent Ginzburg--Landau model}
We consider a diblock copolymer melt containing $N_c$ colloidal particles of core radius $R_0$ with positions and orientations
$\{\mathbf r_i,\mathbf u_i\}_{i=1}^{N_c}$. We describe the diblock copolymer by a conserved order-parameter field $\psi(\mathbf r,t)$. Its dynamics follow the Cahn--Hilliard equation,
\begin{equation}
\frac{\partial\psi}{\partial t}
=M\nabla^2\frac{\delta\mathcal F}{\delta\psi}= M\nabla^2\mu,
\label{eq:chc}
\end{equation}
where $M$ is the mobility and $\mu(\mathbf r,t)=\delta\mathcal F/\delta\psi$ is
the chemical potential. The field noise is set to zero in all simulations.

The physical composition fields $\widetilde\phi_A$ and $\widetilde\phi_B$ are defined only in the melt, outside the colloids. Following Balazs and
coworkers,\cite{balazsMultiScaleModelBinary2000,ginzburgModelingDynamicBehavior2000a} we extend them to continuous fields $\phi_A$ and $\phi_B$ defined throughout the box, related by
\begin{equation}
\widetilde{\phi}_K(\mathbf r)
=\phi_K(\mathbf r)\prod_i\Lambda_c(\mathbf r-\mathbf r_i),
\qquad K=A,B,
\end{equation}
where $\Lambda_c(\mathbf r)=0$ for $|\mathbf r|<R_0$ and $\Lambda_c(\mathbf r)=1$
otherwise. For an AB diblock copolymer with A-block fraction $f=N_A/(N_A+N_B)$, the order parameter is
\begin{equation}
\psi(\mathbf r)=\phi_A(\mathbf r)-\phi_B(\mathbf r)+(1-2f),
\end{equation}
and the corresponding masked field is $\widetilde\psi(\mathbf r)=\psi(\mathbf r)\prod_i\Lambda_c(\mathbf r-\mathbf r_i)$. In the incompressible limit, $\phi_A(\mathbf r)+\phi_B(\mathbf r)=1$. Introducing these fields is numerically convenient and provides a compact framework for the colloid--polymer coupling.

The free energy is 
\begin{equation}
\begin{aligned}
\mathcal{F}[\psi,\{\mathbf r_i,\mathbf u_i\}]
={}&\int \mathrm{d}\mathbf r
\left[
H(\psi) +\frac{D}{2}|\nabla\psi|^2
\right] \\
&+ \frac{B}{2}
\int \mathrm d\mathbf r
\int \mathrm d\mathbf r'\,
G(\mathbf r-\mathbf r')\,
\widetilde{\psi}(\mathbf r)
\widetilde{\psi}(\mathbf r') \\
&+\mathcal F_{cp}[\psi,\{\mathbf r_i,\mathbf u_i\}]
+U_{cc}[\{\mathbf r_i\}].
\end{aligned}
\label{eq:free_energy}
\end{equation}
The first integral contains the local free-energy density and the gradient penalty. The local contribution is
\begin{equation}
H(\psi)
=
\frac{\tau_0}{2}\psi^2
+
\frac{v}{3}(1-2f)\psi^3
+
\frac{u}{4}\psi^4,
\label{eq:local-free-energy}
\end{equation}
whereas the gradient term penalizes spatial variations of $\psi$.

The second term in Eq.~\ref{eq:free_energy} accounts for the connectivity of A and B blocks through the Green's function $G$, defined by
\begin{equation}
\nabla^2 G(\mathbf r-\mathbf r')
=
-\delta(\mathbf r-\mathbf r').
\end{equation} 
The third term in Eq.~\ref{eq:free_energy} couples the particles to the
copolymer field through a hybrid particle--field
scheme,~\cite{balazsMultiScaleModelBinary2000, ginzburgModelingDynamicBehavior2000a,qiHybridParticlecontinuumResolution2016} 
\begin{equation}
\mathcal{F}_{cp}
=
g_{cp}
\sum_{i=1}^{N_c}
\int \mathrm d\mathbf r\,
\psi_c\!\left(|\mathbf r-\mathbf r_i|\right)
\left[
\psi(\mathbf r)-\psi_0(\mathbf r,\mathbf u_i)
\right]^2 .
\label{eq:coupling_F}
\end{equation}

Here, $g_{cp}$ sets the colloid--polymer coupling strength. The preferred composition field of particle $i$ is
\begin{equation}
\psi_0(\mathbf r,\mathbf u_i)
=
\begin{cases}
\psi_b, & \alpha=0,\\[2pt]
\psi_b+(\psi_a-\psi_b)H_s(s_i), & 0<\alpha<1,\\[2pt]
\psi_a, & \alpha=1,
\end{cases}
\label{eq:patch-preference}
\end{equation}
where $\hat{\mathbf r}_i=(\mathbf r-\mathbf r_i)/|\mathbf r-\mathbf r_i|$ is the unit vector from the particle center to the field point, and $s_i \equiv \hat{\mathbf r}_i\cdot\mathbf u_i-\cos(\pi\alpha)$. The smooth step function $H_s(x)=[1+\tanh(k_sx)]/2$ defines the boundary between the two surface regions and $k_s$ controls how rapidly the order parameter changes across the interface. The parameter $\alpha\in[0,1]$ controls the fraction of the surface with preferred composition $\psi_a$; $\alpha=0.5$ corresponds to equal A- and B-selective surface areas.  The envelope function $\psi_c$ localizes the coupling within a shell around the particle and determines its radial profile. We use the compact particle-envelope function
\begin{equation}
\psi_c(r)=
\begin{cases}
\exp\!\left[1-\dfrac{1}{1-\left(r/R_c\right)^{\gamma_{cp}}}\right], & r<R_c,\\[3pt]
0, & r\geq R_c,
\end{cases}
\label{eq:particle-envelope}
\end{equation}
where $R_c=R_0+w$ is the outer cutoff radius of the particle--polymer coupling shell, $R_0$ is the particle-core radius, and $w$ is the fixed shell width. The exponent $\gamma_{cp}$ is chosen such that
$\psi_c(R_0)=1/2$, giving
\begin{equation}
\gamma_{cp}
=
\frac{\ln\left(1+1/\ln 2\right)}
{\ln\left(1+w/R_0\right)}.
\label{eq:gamma-cp}
\end{equation}
Thus, particles with different core radii have the same coupling-shell width but different values of $\gamma_{cp}$. 

The last term in Eq.~\ref{eq:free_energy}, the direct colloid--colloid
interaction $U_{cc}$, is specified in Sec.~\ref{sec:particle-dynamics}. For conciseness and clarity, we summarize all model parameters and definitions in Table~\ref{tab:parameters}.

We should note that, strictly speaking, both the first and the second term in Eq.~\ref{eq:free_energy} should depend on $\widetilde{\psi}$ only, as they describe block copolymer interactions. We express the local and gradient terms as functions of the unmasked $\psi$-field for numerical convenience. The field $\psi$ inside a particle is determined by $\psi_0$ through the particle--field coupling, so as long as the cores do not overlap each other or their periodic images, these terms contribute a constant that is independent of particle position and orientation. This constant cancels in all quantities reported here, since every free-energy difference is taken between states with the same number of particles of the same size. The non-local contribution (second term in Eq.~\ref{eq:free_energy}) has no such property and must be expressed in terms of $\widetilde{\psi}$. The Cahn--Hilliard equation, Eq.~\ref{eq:chc}, is integrated with a semi-implicit pseudospectral scheme, alternating real-space and Fourier-space evaluations, as described in SI Sec.~\ref{app:numerics}.

\subsection{Dynamics of colloidal particles}
\label{sec:particle-dynamics}
The motion of the colloidal particles is described by overdamped Langevin equations. The center-of-mass position $\mathbf r_i$ of particle $i$ evolves as
\begin{equation}
\frac{\mathrm d \mathbf r_i}{\mathrm d t}
=- M_c \nabla_{\mathbf r_i}\left(U_{cc}+\mathcal{F}_{cp}\right)
+\sqrt{2M_ck_BT_c}\,\boldsymbol{\xi}_i(t),
\end{equation}
where $M_c=1/\gamma$ is the translational mobility, $\gamma=6\pi\eta_0R_0$ the friction coefficient and $\eta_0$ the solvent viscosity; $-\nabla_{\mathbf r_i}U_{cc}$ is the force from interparticle interactions and $-\nabla_{\mathbf r_i}\mathcal{F}_{cp}$ the coupling force from the particle--polymer interaction (Eq.~\ref{eq:coupling_F}). The stochastic term is Gaussian white noise with
\begin{equation}
\langle \boldsymbol{\xi}_i(t)\rangle=0,
\qquad
\langle \xi_{i\mu}(t)\xi_{j\nu}(t')\rangle
=\delta_{ij}\delta_{\mu\nu}\delta(t-t').
\end{equation}
The particle noise amplitude is chosen small, $k_BT_c=0.01$ in units of the reported energy scale, so that the Brownian dynamics performs a weakly stochastic search for the free-energy minimum rather than canonical sampling.

Steric interactions between particles are described by the purely repulsive Weeks--Chandler--Andersen potential,
\begin{equation}
U_{\mathrm{WCA}}(r)=
\begin{cases}
4\epsilon
\left[
\left(\dfrac{\sigma_{cc}}{r}\right)^{12}
-\left(\dfrac{\sigma_{cc}}{r}\right)^{6}
\right]+\epsilon,
& r \leq r_{\text{cut}},\\[3pt]
0, & r > r_{\text{cut}},
\end{cases}
\label{eq:wca}
\end{equation}
with $U_{cc}=\sum_{i<j}U_{\mathrm{WCA}}(|\mathbf r_i-\mathbf r_j|)$ and cutoff $r_{\text{cut}}$. We set $\sigma_{cc}=R_c=R_0+w$, where $R_c$ is the outer radius of the particle--polymer coupling envelope. The WCA interaction potential vanishes at $r_{\text{cut}}=2^{1/6}R_c$ and provides a short-range repulsion between particles.

In addition to translational motion, the particles undergo rotational Brownian motion. The orientation $\mathbf u_i$, a unit vector, evolves as
\begin{equation}
\frac{\mathrm d \mathbf u_i}{\mathrm d t}
=
M_r\,\mathbf T_i\times \mathbf u_i
+\sqrt{2M_rk_BT_c}\,\boldsymbol{\xi}_i^{\,r}(t)\times \mathbf u_i,
\end{equation}
with torque
$\mathbf T_i=-\,\mathbf u_i\times\nabla_{\mathbf u_i}\mathcal{F}_{cp}$ (derived in Sec.~\ref{app:torque} of the Supporting Information), rotational mobility $M_r=1/\gamma_r$ and rotational friction $\gamma_r=8\pi\eta_0R_0^3$. The rotational noise satisfies
\begin{equation}
\langle \boldsymbol{\xi}_i^{\,r}(t)\rangle=0,
\qquad
\langle \xi_{i\mu}^{\,r}(t)\xi_{j\nu}^{\,r}(t')\rangle
=
\delta_{ij}\delta_{\mu\nu}\delta(t-t').
\end{equation}
Since the coupling is the only orientation-dependent interaction, the torque vanishes identically for homogeneous particles, which have no orientational degree of freedom in practice. The orientation is updated with Rodrigues' rotation formula and renormalized each step (Sec. \ref{app:rodrigues} of the Supporting Information).

\subsection{Simulation parameters}
\label{sec:simulation-parameters}
All simulations are performed on a simple cubic grid with periodic boundary conditions applied in all three directions, $x$, $y$, and $z$, with a discretization $\Delta h=0.5\,\sigma$. The lateral dimensions $L_x=L_y=L$ are varied as described in Sec.~\ref{sec:finite-size-scaling}; The vertical period is $L_z=20\,\sigma$ for the undoped bulk calculation of Sec.~\ref{sec:bulk-results}, and is varied over $L_z=25$--$40\,\sigma$ in the doped systems so that the thickness dependence of the extracted quantities can be tested explicitly (Sec.~\ref{sec:delta-framework}). The smallest period is excluded from the doped runs because the coupling shell of the largest particles, of diameter $2R_c=2(R_0+w)=23\,\sigma$ at $R_0=9\,\sigma$, would then exceed
$L_z$ and each particle would overlap its own periodic image along the defect line. Because the model uses periodic boundary conditions in $z$, it contains no substrate or free surface. The period $L_z$ therefore represents an infinitely repeated lamellar stack, and the reported formation energies describe bulk dislocation-pair thermodynamics rather than surface-modified thin-film behavior.

Throughout, we use $N_c=2$ colloidal particles, \ie a single pair. Particle types differ only through the preferred composition field in Eq.~\ref{eq:coupling_F}. We set $\psi_a=+1$ and $\psi_b=-1$. Type~A particles have a uniform preference $\psi_0=\psi_a$, whereas Type~B particles have $\psi_0=\psi_b$. For patchy particles, the patch ratio $\alpha$ specifies the fraction of the surface with A-selective preference. Thus, $\alpha=0.5$ gives a balanced Janus particle, while $\alpha=0$ and $\alpha=1$ denote the homogeneous Type~B and Type~A limits, respectively. The particle-core radius is varied over $R_0=2$--$9\,\sigma$. The particle--polymer coupling envelope has the fixed width $w=2.5\,\sigma$, such that $R_c=R_0+w$. Thus, particles of different size have the same coupling-shell width. Particle size is reported below as the reduced diameter $2R_0/\ell_0$.

\begin{table*}[ht]
\centering
\caption{Simulation parameters for the TDGL/Brownian-dynamics simulations of
block copolymer--nanoparticle systems. All quantities are in reduced units,
with $\sigma$ the unit of length and $k_BT$ the unit of energy.}
\label{tab:parameters}
\begin{tabular}{lc|lc}
\hline\hline
\multicolumn{4}{c}{\textbf{Block copolymer}} \\
\hline
Volume fraction & $f = 0.5$ & Quadratic coefficient & $\tau_0 = -0.35$ \\
Cubic coefficient & $v = 1.5$ & Quartic coefficient & $u = 0.5$ \\
Gradient coefficient & $D = 1.0$ & Chain connectivity & $B = 0.008$ \\
Mobility & $M = 1.0$ & Field time step & $\Delta t = 0.01$ \\
Lamellar period\,$^{a}$ & $\lambda_0 \approx 21.3\,\sigma$ & Domain thickness\,$^{a}$ & $\ell_0 = \lambda_0/2 \approx 10.65\,\sigma$ \\
\hline
\multicolumn{4}{c}{\textbf{Particle--polymer coupling}} \\
\hline
Coupling strength & $g_{cp} = 0.5$ & Hard-core radius & $R_0 = 2$--$9\,\sigma$ \\
Coupling shell width & $w = 2.5\,\sigma$ & Coupling cutoff & $R_c = R_0 + w$ \\
Decay parameter & $\gamma_{cp}=\dfrac{\ln(1+1/\ln 2)}{\ln(1+w/R_0)}$ & Range of $\gamma_{cp}$ & $1.10$--$3.64$ \\
Number of particles & $N_c = 2$ & Preferred compositions & $\psi_a=+1$, $\psi_b=-1$ \\
Patch ratio & $\alpha = 0$--$1$ & Patch steepness & $k_s=1$ \\
Mask inside core & $\Lambda_c(r<R_0)=0$ & & \\
\hline
\multicolumn{4}{c}{\textbf{Particle dynamics}} \\
\hline
Fluid viscosity & $\eta_0 = 1.0$ & Brownian noise amplitude & $k_BT_c = 0.01$ \\
Friction & $\gamma = 6\pi\eta_0 R_0$ & Rotational friction & $\gamma_r = 8\pi\eta_0 R_0^3$ \\
Mobility & $M_c = 1/\gamma$ & Rotational mobility & $M_r = 1/\gamma_r$ \\
Particle time step & $\Delta t_c = 0.01$ & & \\
\hline
\multicolumn{4}{c}{\textbf{Interparticle potential}} \\
\hline
WCA strength & $\epsilon = 1$ & WCA diameter & $\sigma_{cc} = R_c$ \\
WCA cutoff & $r_\text{cut} = 2^{1/6}\sigma_{cc}$ & & \\
\hline
\multicolumn{4}{c}{\textbf{System size and protocol}} \\
\hline
Lamellar periods & $N = 5$--$9$ & Lateral box size & $L = 106.5$--$191.5\,\sigma$ \\
Vertical period$^b$ & $L_z = 20$--$40\,\sigma$ & Grid spacing & $\Delta h = 0.5\,\sigma$ \\
Grid dimensions & $213^2$--$383^2 \times 40$--$80$ & Coupling ramp & $t_{\text{ramp}} = 300$ \\
Equilibration (undoped) & $t_{\text{eq}} = 1000$ & Production time & $t_{\text{prod}} = 600$ \\
\hline\hline
\end{tabular}
\begin{flushleft}
\footnotesize $^{a}$Measured from the relaxed defect-free state, not an input parameter. $^{b}$Vertical period $L_z = 20\,\sigma$ is only used for undoped system, and $L_z = 25$--$40\,\sigma$ is used for doped systems.
\end{flushleft}
\end{table*}

\section{Thermodynamic Framework for Particle--Defect Interactions}
\subsection{Reference states and formation energies}

The defect-free reference state is obtained by relaxing an initial stripe pattern until the free energy converges. The defect-containing reference state is prepared from a two-dimensional stripe pattern containing one dislocation pair. The pattern is normalized to the order-parameter range, extended uniformly along $z$, and relaxed using Eq.~\ref{eq:chc}. Because the field noise amplitude is zero, the relaxation is deterministic. The initial pattern selects the basin of attraction, whereas the core separation, and surrounding lamellar distortion follow from free-energy minimization. We note that the defect-containing state breaks the symmetry with respect to exchanging $\phi_A$ and $\phi_B$: A-lamellae branch and B-lamellae end at the defect cores (see Fig. \ref{fig:sketch}).

Dislocation cores are identified in the mid-plane order-parameter field from the determinant of the two-dimensional Hessian, $\det \mathcal H(\psi)=\psi_{xx}\psi_{yy}-\psi_{xy}^{2}$. Candidate pixels are selected by a strongly negative Hessian determinant and small $|\psi|$. Connected candidate regions define core clusters, and the centers of the two strongest clusters are used as the dislocation-core positions.

Particles are inserted either at the two core positions, in the far-field lamellar region of the defect-containing state, or at random locations in the defect-free state. To avoid insertion transients, the particle--polymer coupling strength is increased from zero to $g_{cp}=0.5$ during a warm-up stage. The system is then evolved at fixed $g_{cp}=0.5$. Particle positions and orientations are evolved by Brownian dynamics at $k_BT_c=0.01$. Reported free energies are averaged over the production stage.

We compare four relaxed states for a system with $N$ lamellar periods and lateral dimensions $L_x=L_y=L$: the defect-free and defect-containing states, with free energies $F_l(L,N)$ and $F_d(L,N)$, respectively, and the corresponding states containing two particles, with free energies $F_{lp}(L,N,\mathbf R,R_0)$ and $F_{dp}(L,N,\mathbf R,R_0)$. Here, $\mathbf R=\{\mathbf r_1,\mathbf r_2\}$ denotes the particle positions and $R_0$ is the particle-core radius. For patchy particles, the free energies also depend on the particle orientations, which are minimized together with $\mathbf R$.

The formation energy of a dislocation pair in the undoped system is
\begin{equation}
\Delta F_d(N)
=
F_d(L_l^*,N)-F_l(L_l^*,N),
\label{eq:dFd}
\end{equation}
where $L_l^*$ is the commensurate lateral box size that minimizes the free-energy density of the defect-free lamellar state at fixed $N$. The corresponding formation energy in the presence of two particles is
\begin{equation}
\Delta F_{dp}(N)
=
F_{dp}(L_l^*,N,\mathbf R_{\mathrm{core}}^*,R_0)
-
F_{lp}(L_l^*,N,\mathbf R_{\mathrm{lam}}^*,R_0).
\label{eq:dFdp}
\end{equation}
Here, $\mathbf R_{\mathrm{core}}^*$ and
$\mathbf R_{\mathrm{lam}}^*$ are the particle configurations that minimize the free energy in the defect-containing and defect-free states, respectively.

Equation~\eqref{eq:dFd} is the free-energy cost of a dislocation pair in the undoped system. It includes the core contribution and the elastic distortion around the pair.\cite{andersonTheoryDislocations2017} Equation~\eqref{eq:dFdp} defines the same quantity at fixed particle number. Since the two reference states contain the same number of particles, their difference excludes the single-particle solvation contribution and isolates the effect of the particles on the defect formation energy. Table~\ref{tab:thermo-notation} summarizes the notation.
\begin{table}[t]
\caption{Notation used in the thermodynamic framework. Subscripts $l$ and $d$
denote the defect-free and defect-containing states, $p$ the presence of two particles.}
\label{tab:thermo-notation}
\begin{ruledtabular}
\begin{tabular}{p{0.22\columnwidth}p{0.70\columnwidth}}
Symbol & Definition \\
\hline
$N$, $L$, $L_z$ & Lamellar periods ($2N$ domains), lateral box size, vertical period. \\
$L_l^*(N)$ & Commensurate $L$ minimizing the free-energy density of the defect-free state; $L_l^*\approx\lambda_0N$. \\
$\lambda_0$, $\ell_0$ & Lamellar period and domain thickness, $\ell_0=\lambda_0/2$. \\
$R_0$, $\alpha$ & Particle-core radius (reduced diameter $2R_0/\ell_0$) and patch ratio; $\alpha=0,1$ give Type~B, Type~A and $\alpha=0.5$ the balanced Janus particle. \\
$\mathbf R^*_{\mathrm{lam}}$, $\mathbf R^*_{\mathrm{core}}$, $\mathbf R^*_{\mathrm{far}}$ & Relaxed particle positions (with orientations, for patchy particles): in the defect-free lamellae, at the dislocation cores, and in the far-field lamellae of the defect-containing state. \\
$F_{l}$, $F_{d}$, $F_{lp}$, $F_{dp}$ & Free energies of the four relaxed states, at arguments $(L,N)$ and, with particles, $(\mathbf R,R_0)$. \\
$\Delta F_d$, $\Delta F_{dp}$ & Dislocation-pair formation energy, undoped and with two particles at the cores, Eqs.~\eqref{eq:dFd},~\eqref{eq:dFdp}. \\
$\Delta F_{dp}^{\mathrm{far}}$ & As $\Delta F_{dp}$, particles held in the far-field lamellae, Eq.~\eqref{eq:dFdp-far}. \\
$\Delta F_{\mathrm{bind}}$ & Particle--core binding free energy, Eq.~\eqref{eq:dFbind}; negative indicates binding. \\
$\Delta F^\infty$ & Thermodynamic-limit ($L\rightarrow\infty$) value of any $\Delta F$, Eq.~\eqref{eq:fss-general}. \\
$\delta(R_0)$ & Particle-induced shift, $\Delta F_{dp}^\infty-\Delta F_d^\infty$, Eq.~\eqref{eq:delta}; negative indicates stabilization. \\
$c$, $L_z^*$ & Formation energy per unit length, $\Delta F_d^\infty=cL_z$; and $L_z^*=|\delta|/c$, Eq.~\eqref{eq:Lz-star}. \\
$n_p$ & Line density of particle pairs along the dislocation, Eq.~\eqref{eq:line-density}. \\
\end{tabular}
\end{ruledtabular}
\end{table}

\subsection{Binding free energy of particles to the dislocation core}
\label{sec:binding-framework}

The formation energies in Eqs.~\eqref{eq:dFd} and~\eqref{eq:dFdp} quantify the effect of particles on the formation energy of a dislocation pair, but they do not quantify the particles' preference for the defect core. We therefore define the particle--core binding free energy by comparing two particle configurations within the same defect-containing state.

In the core configuration, the particles occupy the two dislocation cores at positions $\mathbf R_{\mathrm{core}}^*$. In the far-field configuration, the particles are placed in the undistorted lamellar region at $\mathbf R_{\mathrm{far}}^*$. Both configurations have identical $L$, $N$, $R_0$, and surface-patch parameters. The binding free energy is
\begin{equation}
\Delta F_{\mathrm{bind}}(N)
=
F_{dp}(L_l^*,N,\mathbf R_{\mathrm{core}}^*,R_0)
-
F_{dp}(L_l^*,N,\mathbf R_{\mathrm{far}}^*,R_0).
\label{eq:dFbind}
\end{equation}
A negative value of $\Delta F_{\mathrm{bind}}$ indicates that the particles prefer the dislocation cores. A positive value indicates that the particles prefer the far-field lamellar environment.

For the far-field reference, the particle centers are constrained to remain in the far-field lamellar region while the polymer field and, for patchy particles, the particle orientations relax. Without this constraint, particles with a negative binding free energy would migrate to the cores and the far-field state would not be defined.
\subsection{Finite-size scaling}
\label{sec:finite-size-scaling}
The formation energies defined above depend on the lateral box size $L$. At the commensurate box size, the defect-free lamellar state is free of the elastic mismatch associated with an incommensurate lateral period. The defect-containing state, however, contains a long-range strain field that interacts with periodic images of the dislocation pair. Consequently, $F_d(L,N)$ and the corresponding formation energy $\Delta F_d(L)$ retain a dependence on $L$. The same applies to the doped formation energy $\Delta F_{dp}(L)$ and the binding free energy $\Delta F_{\mathrm{bind}}(L)$. We therefore extrapolate all three quantities to the limit $L\rightarrow\infty$.

In substrate-templated or laterally confined systems, the guiding pattern or channel fixes the lamellar spacing and can screen the defect-induced strain field. The defect free energy then approaches a size-independent value once the system spans several lamellar periods.\cite{nagpalFreeEnergyDefects2012a,takahashiDefectivityLaterallyConfined2012a} The present system is untemplated. Although the opposite Burgers vectors of the dislocation pair screen the far-field distortion,\cite{hurMolecularPathwaysDefect2015a} the pair still interacts with its periodic images. We therefore perform an explicit finite-size extrapolation.

For each number of lamellar periods $N$, we determine the commensurate lateral lamellar state with respect to $L$ by minimizing the free-energy density of the defect-free lamellar state. The resulting sizes satisfy $L_l^*(N)\approx 21.3 N$, corresponding to a lamellar period $\lambda_0\approx 21.3\,\sigma$ and a domain thickness $\ell_0=\lambda_0/2\approx 10.65\,\sigma$.

Over the range of box sizes studied, the formation energies are described by
\begin{equation}
\Delta F(L)
=
\Delta F^{\infty}
+
\frac{A}{L^2},
\label{eq:fss-general}
\end{equation}
where $\Delta F^{\infty}$ is the thermodynamic-limit value and $A$ describes the leading finite-size correction. We fit the sequences
$\{\Delta F_d(L_l^*(N))\}$, $\{\Delta F_{dp}(L_l^*(N))\}$, and $\{\Delta F_{\mathrm{bind}}(L_l^*(N))\}$ separately to Eq.~\eqref{eq:fss-general}.

The quantities $\Delta F_d^{\infty}$ and $\Delta F_{dp}^{\infty}$ are the formation energies of an isolated dislocation pair and an isolated particle--defect complex, respectively. Similarly, $\Delta F_{\mathrm{bind}}^{\infty}$ is the binding free energy of particles to an isolated dislocation pair. A positive formation energy indicates that the defect-free lamellar state is favored in the dilute-defect limit. A negative value indicates that the defect-containing state is favored. These quantities do not determine the defect density at finite defect concentration, where defect--defect interactions and configurational entropy also contribute.

\subsection{The particle-induced shift and its thickness scaling}
\label{sec:delta-framework}
In the periodic geometry considered here, the dislocation pair is a line defect that is uniform along $z$. Its formation energy is therefore proportional to the vertical period,
\begin{equation}
\Delta F_d^{\infty}(L_z)=cL_z,
\label{eq:extensive}
\end{equation}
where $c$ is the formation energy per unit length of the dislocation pair. In
contrast, the free-energy contribution of a particle pair is localized near the
defect core and does not scale with $L_z$. We therefore define the
particle-induced shift
\begin{equation}
\delta(R_0)
\equiv
\Delta F_{dp}^{\infty}
-
\Delta F_d^{\infty}.
\label{eq:delta}
\end{equation}
A negative value of $\delta$ indicates that the particles lower the dislocation
formation energy, whereas a positive value indicates destabilization. We test
the independence of $\delta$ from $L_z$ in
Sec.~\ref{sec:homogeneous-results} and use it as the measure of the local
particle-induced change in defect thermodynamics.

Because $\delta$ is localized whereas $\Delta F_d^{\infty}$ grows linearly with
$L_z$, the doped formation energy,
$\Delta F_{dp}^{\infty}=cL_z+\delta$, can become negative only for vertical system sizes $L_z$ below
\begin{equation}
L_z^*(R_0)=\frac{|\delta(R_0)|}{c},
\qquad
\delta<0.
\label{eq:Lz-star}
\end{equation}
For larger $L_z$, the defect-free state remains thermodynamically preferred at
the fixed particle-pair content considered here. More generally, if independent particle pairs decorate the dislocation with line density $n_p$, the defect-containing state is favored when
\begin{equation}
n_p|\delta|>c.
\label{eq:line-density}
\end{equation}
Thus, defect stabilization depends on the density of particle pairs decorating
the dislocation, rather than on the properties of an isolated particle.

\section{Results}
\subsection{Bulk defect formation energy}
\label{sec:bulk-results}
\begin{figure}[t!]
    \centering
\includegraphics[width=0.95\linewidth]{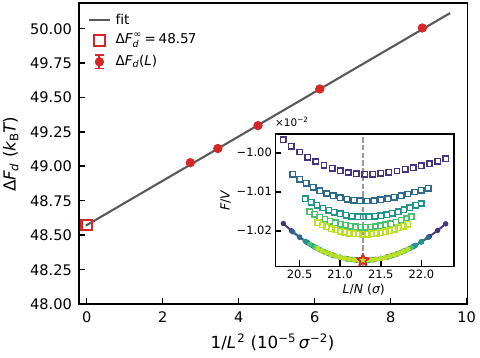}
\caption{Defect formation energy $\Delta F_{d}(L)$ versus $1/L^{2}$ in an undoped lamellar block copolymer with $L_z=20\,\sigma$. The solid line shows a fit to Eq.~\ref{eq:fss-general}, yielding the thermodynamic-limit value $\Delta F_{d}^{\infty}=48.57\pm\, 0.01 k_BT$ (empty square). Inset: free-energy density $F/V$ of the defect-free lamellar (circles) and defect-containing lamellar (squares) state as a function of $L/N$ for different lamellar periods $N\in[5,6,7,8,9]$ (dark blue to bright yellow). The asterisk marks the commensurability minimum used to determine the optimal lateral box size $L_{\text{l}}^*$ for the defect-free lamellar state.}
\label{fig:bulk_opti_finitesize_scaling}
\end{figure}

We first quantify the thermodynamic cost of creating a dislocation pair in the absence of dopant particles. Starting from representative order-parameter fields for the defect-free and defect-containing lamellar states with smallest period $N=5$ shown in Fig.~\ref{fig:sketch}, we optimize the lateral box size $L$ for individual $N=5,6,7,8,9$ by minimizing $F/V$ with respect to $L/N$. The inset of Fig.~\ref{fig:bulk_opti_finitesize_scaling} shows the resulting commensurability minima for both defect-free and defect-containing lamellar states; the minimum of the defect-free state defines $L^{*}_{l}$ for each $N$. Evaluating $\Delta F_{d}=F_{d}(L^{*}_{l})-F_{l}(L^{*}_{l})$ at these optimal sizes gives the finite-size sequence plotted in Fig.~\ref{fig:bulk_opti_finitesize_scaling}, which increases linearly with $1/L^{2}$ and is well described by Eq.~\eqref{eq:fss-general}, consistent with periodic-image interactions as the dominant source of size dependence. Extrapolation to the thermodynamic limit gives $\Delta F_{d}^{\infty}=48.57\pm0.01 \,k_BT$, the reference formation energy of the undoped lamellar phase against which the effect of dopant particles is quantified in the following sections.
\subsection{Defect stabilization by homogeneous particles}
\label{sec:homogeneous-results}
\begin{figure*}[t]
    \centering
\includegraphics[width=\linewidth]{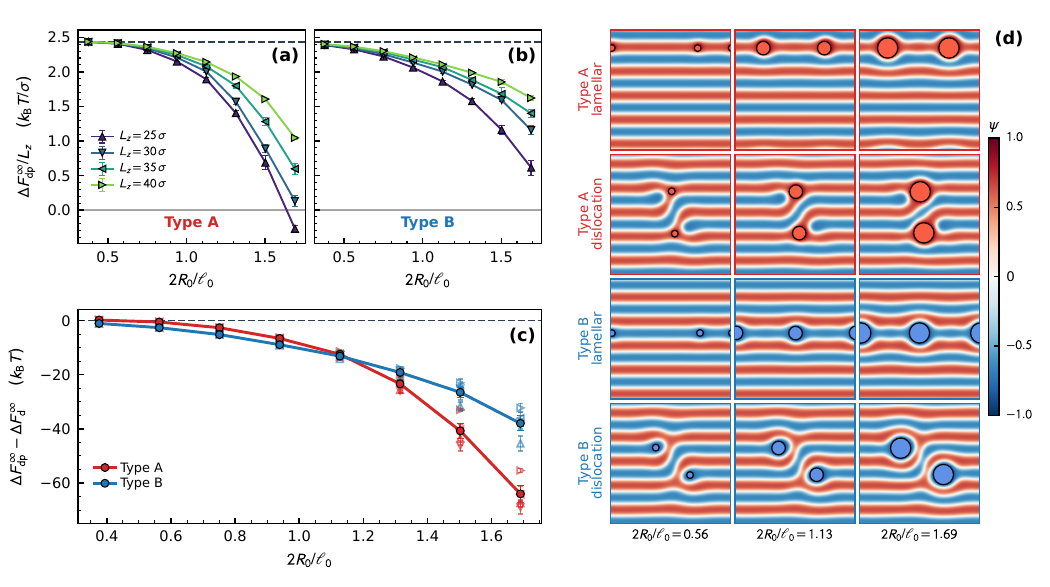}
\caption{Effect of homogeneous colloidal particle pairs on the
dislocation-pair formation energy. (a,b) Thermodynamic-limit formation energy per unit thickness, $\Delta F_{dp}^{\infty}/L_z$, as a function of reduced particle diameter $2R_0/\ell_0$ for (a) Type~A and (b) Type~B particles, at $L_z=25$--$40\,\sigma$. The dashed line denotes the undoped reference, $\Delta F_d^\infty/L_z=c=2.43\,k_BT/\sigma$. (c) Particle-pair-induced shift, $\delta=\Delta F_{dp}^{\infty}-\Delta F_d^{\infty}$. Filled symbols and solid lines show averages over $L_z$; error bars denote the standard error across thicknesses, faint open symbols show the individual thicknesses, $\vartriangle\,25\,\sigma$, $\triangledown\,30\,\sigma$,
$\vartriangleleft\,35\,\sigma$, and $\vartriangleright\,40\,\sigma$. (d) Mid-plane order-parameter fields $\psi(\mathbf r)$ at $L_z=25\,\sigma$ and $N=5$ for $2R_0/\ell_0=0.56$, $1.13$, and $1.69$. From top to bottom, the rows show defect-free and defect-containing states with Type~A particles, followed by defect-free and defect-containing states with Type~B particles. Particle colors identify the particle type and are unrelated to the $\psi$ color scale.}
\label{fig3:homo-particles}
\end{figure*}

We next examine how a pair of homogeneous colloidal particles modifies the formation energy of a dislocation pair. Type~A particles prefer A-rich domains,
whereas Type~B particles prefer B-rich domains. The particle size is reported as the reduced diameter $2R_0/\ell_0$, where $\ell_0=10.65\,\sigma$ is the equilibrium domain thickness. Particles with $2R_0/\ell_0<1$ fit within a single domain, whereas larger particles span an internal interface and distort the surrounding lamellae.

The undoped formation energy is proportional to the vertical period, $\Delta F_d^\infty=cL_z$, with $c=2.43\,k_BT/\sigma$. This behavior follows because the dislocation pair is a line defect that is uniform along $z$. We therefore plot the doped formation energy per unit thickness in Fig.~\ref{fig3:homo-particles}(a,b). As $2R_0/\ell_0\rightarrow0$,
$\Delta F_{dp}^\infty/L_z$ approaches the undoped value $c$. For larger particles, $\Delta F_{dp}^\infty/L_z$ decreases for both particle types. The reduction is larger at smaller $L_z$, as expected for a localized contribution
to an extensive line-defect energy.

Figure~\ref{fig3:homo-particles}(c) shows the particle-pair-induced shift, $\delta$ (Eq.~\eqref{eq:delta}) which isolates the particle contribution. For
$2R_0/\ell_0\lesssim 1$, $\delta$ is independent of the vertical period over $L_z=25$--$40\,\sigma$. The particle-induced change is therefore local to the particle--defect complex over this size range. At the two largest particle sizes a residual $L_z$ dependence remains. Although $L_z$ exceeds the coupling-shell diameter $2R_c=2(R_0+w)$ in all simulations, the margin is small at these sizes ($2\sigma$), and the particle-induced distortion still overlaps with that of its own periodic image along the defect line. We do not extrapolate $\delta$ to $L_z\rightarrow\infty$: the in-plane scaling of Eq.~\eqref{eq:fss-general} follows from the dipolar elastic interaction between the dislocation pair and its lateral images; In contrast, the residual $L_z$ dependence originates from the overlap of a localized particle perturbation with its own images along $z$, for which no corresponding functional form is available. We therefore quote $\delta$ at these sizes as a mean over thicknesses with the spread as a systematic uncertainty. This does
not affect the conclusions drawn here, which rely on the sign of $\delta$ and its dependence on particle size and surface chemistry rather than its precise value at any single size.

The magnitude of $\delta$ increases with particle size for both surface preferences. Below $2R_0/\ell_0\approx1$, Type~A and Type~B particles give similar shifts with differences of only a few $k_BT$: a particle that fits within one domain perturbs the defect-free lamellae and the defect-containing state almost equally. Above the domain thickness the two types separate, and at $2R_0/\ell_0=1.69$ Type~A particles give $\delta=-63.8\pm3.0\,k_BT$, whereas Type~B particles give $-37.9\pm2.8\,k_BT$. The A-selective surface is therefore more favorable in the local environment of the dislocation core for dislocation defects involving branching A-lamellae.

The order-parameter fields in Fig.~\ref{fig3:homo-particles}(d) illustrate the origin of this stabilization. In the defect-free state, particles larger than a domain displace the surrounding lamellae. Near a dislocation core, the same particles occupy a region that is already distorted. They replace strained polymer rather than introduce a comparable distortion into an otherwise uniform lamellar environment. The resulting free-energy reduction increases with particle size, which leads to the decreasing trend in $\delta$. Homogeneous particles therefore stabilize dislocation pairs when they are large enough to couple strongly to the defect-core structure. The magnitude of the effect depends on particle size and surface preference, whereas its ability to reverse the sign of the formation energy additionally depends on the density of particle pairs along the dislocation.

\subsection{Defect Destabilization by Janus particles}
\label{sec:janus-results}
\begin{figure}[h]
    \centering
\includegraphics[width=\linewidth]{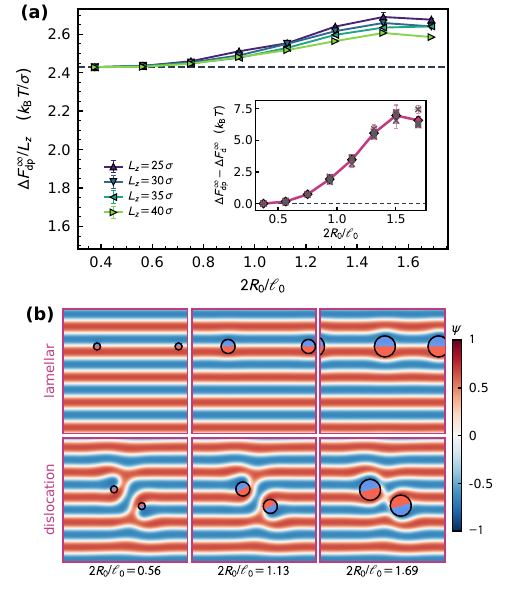}
\caption{Effect of Janus colloids ($\alpha=0.5$) on the
dislocation-pair formation energy. (a)~Thermodynamic-limit formation energy per unit thickness, $\Delta F_{dp}^{\infty}/L_z$, as a function of reduced particle size $2R_0/\ell_0$, at vertical periods $L_z=25$--$40\,\sigma$. The dashed line
marks the undoped reference $\Delta F_{d}^{\infty}/L_z=c=2.43\,k_BT/\sigma$.
Inset: the particle-induced shift $\delta=\Delta F_{dp}^{\infty}-\Delta F_{d}^{\infty}$. Filled symbols and solid lines show averages over $L_z$; error bars denote the standard error across thicknesses, faint open symbols show the individual thicknesses, $\vartriangle\,25\,\sigma$, $\triangledown\,30\,\sigma$,
$\vartriangleleft\,35\,\sigma$, and $\vartriangleright\,40\,\sigma$. (b)~Mid-plane order-parameter fields $\psi(\mathbf{r})$ in the final production frame at $L_z=25\,\sigma$ and $N=5$ (box $106.5\times106.5\times25$) for $2R_0/\ell_0=0.56$, $1.13$ and $1.69$, for the defect-free lamellar state (upper row) and the defect-containing state (lower row). The two-tone particle fill indicates the A- and B-selective faces and the short bar marks the in-plane projection of the patch axis; both are projections of a three-dimensional patch onto the mid-plane and are unrelated to the $\psi$ color scale.}
\label{fig4:patchy-particles}
\end{figure}

We next consider Janus particles with equal A- and B-selective surface coverage ($\alpha=0.5$). Figure~\ref{fig4:patchy-particles}(a) shows that these particles affect the dislocation pair in a qualitatively different way from the homogeneous particles of Sec.~\ref{sec:homogeneous-results}. The formation energy per unit thickness lies above the undoped reference at every size and for every thickness studied, and the particle-induced shift $\delta$ (inset) is positive throughout: the Janus particle enhances the cost of the dislocation pair rather than lowering it. 
The magnitude of the destabilization is modest but systematic. Below $2R_0/\ell_0\approx1$, $\delta$ is within 2.5~$k_BT$: a particle smaller than one lamellar domain neither stabilizes nor destabilizes the defect, just as for the homogeneous types. Above this size $\delta$ rises steadily, reaching a maximum of $\approx7\,k_BT$ near $2R_0/\ell_0\approx1.5$. The contrast with the homogeneous case is in both sign and scale: over the same size range, homogeneous particles lower $\delta$ by several tens of $k_BT$ [Fig.~\ref{fig3:homo-particles}(c)], whereas the balanced Janus particle raises it by a few.

At the largest particle size, $\delta$ decreases slightly. At this size, the particle is comparable to the dislocation-core region and perturbs the strained morphology. The two dislocation cores can then move closer together, reducing the free energy of the defect-containing state and hence $\delta$. The reduced core separation is visible in the morphologies at $2R_0/\ell_0=1.69$ in Fig.~\ref{fig4:patchy-particles}(b). The dislocation pair does not annihilate yet. The measured $\delta$ remains positive, whereas annihilation would drive $\Delta F_{dp}^{\infty}$ toward zero and thus give $\delta\rightarrow-cL_z$.
\begin{figure*}[t]
    \centering
\includegraphics[width=\linewidth]{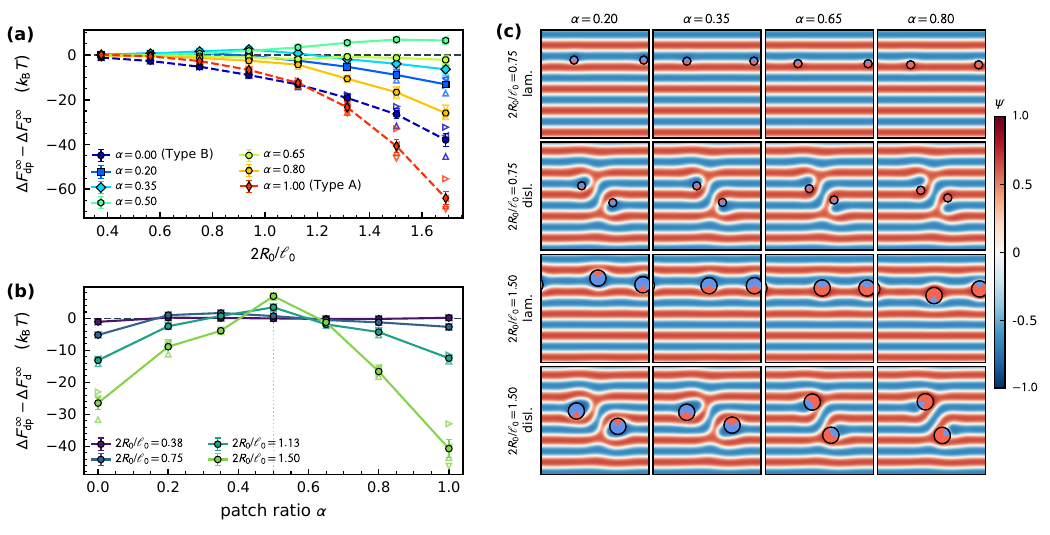}
\caption{Effect of the patch ratio $\alpha$ on the particle-induced shift of the dislocation-pair formation energy, $\delta=\Delta F_{dp}^{\infty}-\Delta F_{d}^{\infty}$. (a)~$\delta$ as a function of reduced particle size $2R_0/\ell_0$ at fixed patch ratio, for $\alpha=0.20$--$0.80$; the endpoints $\alpha=0$ and $\alpha=1$ (dashed) are the homogeneous Type~B and Type~A particles of Fig.~\ref{fig3:homo-particles}, obtained from the same simulations.
(b)~$\delta$ against $\alpha$ at fixed particle size; the dotted vertical line marks balanced coverage, $\alpha=0.5$. In both panels filled symbols are means over $L_z=25$--$40\,\sigma$ with error bars giving the standard error across thicknesses, and faint open symbols show the individual thicknesses, $\vartriangle\,25\,\sigma$, $\triangledown\,30\,\sigma$,
$\vartriangleleft\,35\,\sigma$, and $\vartriangleright\,40\,\sigma$. (c)~Mid-plane order-parameter fields $\psi(\mathbf{r})$ in the final production frame at $L_z=30\,\sigma$ and $N=5$ (box $106.5\times106.5\times30$), for $\alpha=0.20$, $0.35$, $0.65$ and $0.80$. Rows show, from top to bottom, the defect-free lamellar and defect-containing states at $2R_0/\ell_0=0.75$, and the same two states at $1.50$. }
\label{fig6:patch_ratio}
\end{figure*}

The morphologies in Fig.~\ref{fig4:patchy-particles}(b) explain why balanced Janus particles behave differently from homogeneous particles. A homogeneous particle has one preferred composition and can reside within a corresponding domain. A balanced Janus particle must place its A-selective and B-selective faces in A-rich and B-rich regions, respectively. It therefore preferentially occupies an internal interface with a fixed orientation. This condition is readily satisfied in defect-free lamellae, where the interfaces are flat and parallel. Near a dislocation core, however, the interfaces are curved and their orientation varies over the particle surface. Both surface preferences cannot be satisfied simultaneously, which raises the free energy of the defect-containing state relative to the defect-free state.

The binding free energy supports this interpretation. For balanced Janus particles, $\Delta F_{\mathrm{bind}}^\infty$ remains close to zero over most of the size range and becomes positive for the largest particles (Fig.~\ref{fig7:binding}(a)). Thus, these particles are initially indifferent to the core and are eventually expelled from it, unlike homogeneous particles, which bind preferentially to the cores. A particle that does not gain free energy from occupying the core cannot lower the cost of forming it. The same orientational coupling also breaks the symmetry in $z$: starting from the mid-plane, homogeneous particles stay approximately coplanar at the two cores, whereas Janus particles separate along the defect line(see SI Fig.~\ref{SI_fig:separation}).
 
\subsection{Tuning defect stability with the patch ratio}
\label{sec:patch-ratio-results}
The balanced Janus particle of Sec.~\ref{sec:janus-results} and the homogeneous particles of Sec.~\ref{sec:homogeneous-results} have opposite effects on the dislocation-pair formation energy. We therefore vary the patch ratio $\alpha$, defined as the fraction of the particle surface with A-selective preference, so that $\alpha=0$ and $\alpha=1$ denote the homogeneous Type~B and Type~A particles. Figure~\ref{fig6:patch_ratio}(a) shows $\delta$ as a function of particle size at fixed $\alpha$. Particles smaller than one domain sit entirely within a single domain, so their surface pattern has almost no effect on the formation energy. The curves separate only for larger particles, spanning more than $69\,k_BT$ at $2R_0/\ell_0\approx1.69$. There, $\alpha=0.50$ remains slightly above zero, whereas $\alpha=0.20$, $0.35$, $0.65$ and $0.80$ become increasingly negative, approaching the homogeneous limits.

Figure~\ref{fig6:patch_ratio}(b) shows the dependence on $\alpha$ at fixed particle size. The curve is single-peaked, with a maximum near balanced coverage and minima at the homogeneous limits, and its amplitude grows with particle size. This behavior follows from the orientational frustration described in Sec.~\ref{sec:janus-results}: a near-balanced particle cannot satisfy both surface preferences in the curved environment of a core, therefore it gains nothing by staying close to the defect. As $\alpha$ moves away from the balanced value, the minority patch becomes smaller and the orientational constraint is relaxed. The particle then behaves increasingly as a homogeneous inclusion and lowers $\delta$ by replacing strained polymer near the core. The maximum of $\delta$ therefore identifies the strongest orientational frustration, while the two wings describe the crossover to homogeneous-particle behavior.

An asymmetry between the two branches appears only for the largest particles. This asymmetry reflects the polarity of the dislocation core, not an asymmetry of the free-energy functional. The two dislocation core polarities are not equivalent at fixed initial configuration: one terminating domain forms a convex end cap, whereas the neighboring domain wraps around it with opposite curvature. A particle with unequal A- and B-selective surface coverage couples to this polarity. Thus, $\delta(\alpha)$ need not equal $\delta(1-\alpha)$ for a fixed core configuration. This effect appears only when the particle is large enough to resolve the curvature of the core. The morphologies in Fig.~\ref{fig6:patch_ratio}(c) are consistent with this interpretation. At $2R_0/\ell_0=0.75$, the lamellae remain nearly unchanged for all values of $\alpha$ in both defect-free and defect-containing states. At $2R_0/\ell_0=1.50$, the particle reorganizes the nearby domains. Its position and orientation relative to the dislocation core vary with patch ratio, and the
defect-containing morphologies differ accordingly.

\subsection{Binding to the core versus stabilization of the defect}
\label{sec:binding-results}

\begin{figure}[t]
    \centering
\includegraphics[width=\linewidth]{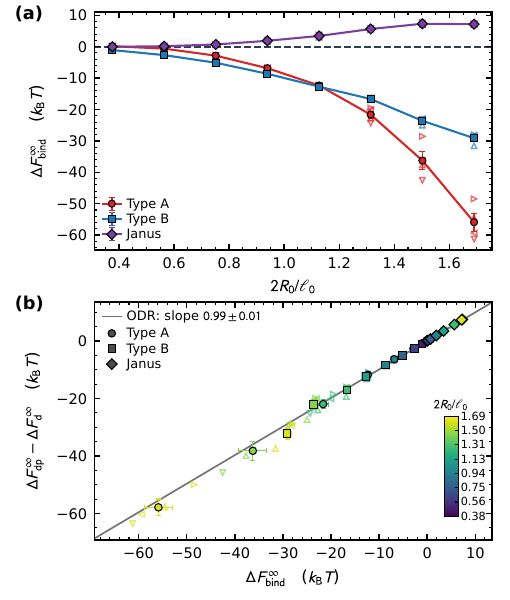}
\caption{Binding free energy of the particle pair to the dislocation cores and its relation to the defect formation energy. (a)~Binding free energy
$\Delta F_{\mathrm{bind}}^{\infty}$, Eq.~\eqref{eq:dFbind}, versus reduced particle size $2R_0/\ell_0$ for Type~A (red), Type~B (blue), and balanced Janus ($\alpha=0.5$, purple) particles. (b)~Particle-pair-induced shift $\delta$ versus $\Delta F_{\mathrm{bind}}^{\infty}$, with color denoting particle size. The grey line is an orthogonal-distance regression (ODR) with slope $1.00\pm0.01$ and intercept $0.04\,k_\mathrm{B}T$. In both panels, filled symbols are means over $L_z=25$--$40\,\sigma$ with error bars showing the standard error across thicknesses, faint open symbols show the individual thicknesses, $\vartriangle\,25\,\sigma$, $\triangledown\,30\,\sigma$,
$\vartriangleleft\,35\,\sigma$, and $\vartriangleright\,40\,\sigma$.}

\label{fig7:binding}
\end{figure}
The three particle types differ qualitatively in their affinity for the dislocation cores [Fig.~\ref{fig7:binding}(a)]. Homogeneous particles bind increasingly strongly with increasing size: $\Delta F_{\mathrm{bind}}^{\infty}$ is within a few $k_BT$ of zero for the smallest particles and decreases steeply above $2R_0/\ell_0\approx1.1$. Balanced Janus particles remain near zero over the small-particle range and become positive above $2R_0/\ell_0\approx1.3$. Thus, large balanced Janus particles are expelled from the cores rather than merely indifferent to them. This behavior explains the destabilization reported in Sec.~\ref{sec:janus-results}. Homogeneous particles lower the formation energy because they prefer the distorted cores to the undistorted lamellae, and this preference increases with particle size. Orientational frustration removes this driving force for balanced Janus particles.

To relate the core binding to defect stabilization, we calculate the formation energy for particles held in the far-field lamellae,
\begin{equation}
\Delta F_{dp}^{\mathrm{far}}(L)
=
F_{dp}(L,N,\mathbf R_{\mathrm{far}}^*,R_0)
-
F_{lp}(L,N,\mathbf R_{\mathrm{lam}}^*,R_0).
\label{eq:dFdp-far}
\end{equation}
Equations~\eqref{eq:dFdp},~\eqref{eq:dFbind} and~\eqref{eq:dFdp-far} then give the identity
\begin{equation}
\Delta F_{dp}(L)
=
\Delta F_{dp}^{\mathrm{far}}(L)
+
\Delta F_{\mathrm{bind}}(L),
\label{eq:decomposition}
\end{equation}
which holds at every box size. Extrapolating each term separately with
Eq.~\eqref{eq:fss-general} and subtracting $\Delta F_d^{\infty}$ yields
\begin{equation}
\delta
=
\left[
\Delta F_{dp}^{\mathrm{far},\infty}-\Delta F_d^{\infty}
\right]
+
\Delta F_{\mathrm{bind}}^{\infty}.
\label{eq:delta-decomposition}
\end{equation}
The bracketed term measures how unequally the far-field particles perturb the defect-containing and defect-free reference states. It vanishes when the two are
perturbed equally, in which case
\begin{equation}
\delta \simeq \Delta F_{\mathrm{bind}}^{\infty}.
\label{eq:slope-one}
\end{equation}

The data follow this relation over the range studied
(Fig.~\ref{fig7:binding}(b)). The orthogonal-distance regression gives a slope of $1.00\pm0.01$ and an intercept of $0.04\,k_BT$. Thus, within the uncertainty of the calculations, the far-field particles affect the defect-containing and defect-free reference states equally. In this parameter range, the particle--core binding free energy provides a quantitative measure of the particle-induced shift in dislocation formation energy. 

\section{Conclusion}
\label{sec:con}
We investigated how colloidal dopants modify the formation energy of dislocation pairs in lamellar block copolymer systems using a hybrid time-dependent Ginzburg--Landau and Brownian-dynamics model. Free-energy minimization and finite-size scaling yielded thermodynamic-limit formation energies and resolved the effects of particle size and surface patterning.

The particle-pair-induced shift, $\delta=\Delta F_{dp}^{\infty}-\Delta F_d^{\infty}$, separates the localized particle contribution from the line energy of the dislocation pair. The undoped formation energy scales as $cL_z$, whereas $\delta$ is independent of $L_z$ for particles up to approximately one domain diameter. Independent particle pairs therefore stabilize a dislocation when $n_p|\delta|>c$, where $n_p$ is their line density along the defect line.

Homogeneous particles lower the formation energy increasingly strongly with size. Particles smaller than one domain change the formation energy by only a few $k_BT$. Larger particles replace strained polymer near the dislocation cores and lower the energy by several tens of $k_BT$. At $2R_0/\ell_0=1.69$, Type~A and Type~B particles give $\delta=-63.8\pm3.0\,k_BT$ and $\delta=-37.9\pm2.8\,k_BT$, respectively. The A-selective surface is therefore more favorable in the local environment of the dislocation cores.

Balanced Janus particles instead raise the formation energy by a few $k_BT$. Their A- and B-selective faces can be accommodated at the flat interfaces of the defect-free lamellae but not near the curved dislocation cores. Varying the patch ratio interpolates between these limits. Surface patterning has little effect on the defect formation energy for particles smaller than one domain, whereas large particles can shift it by more than $69\,k_BT$, from a positive shift near balanced coverage to strongly negative shifts for homogeneous particles. The response becomes asymmetric about $\alpha=0.5$ for large particles, which
we attribute to coupling between unequal surface coverage and the polarity of
the dislocation core.

The particle--core binding free energy provides an independent measure of the same effect. Across all particle types, the relation $\delta=\Delta F_{\mathrm{bind}}^\infty$ holds with a fitted slope of $1.00\pm0.01$. Thus, within the parameter range studied, the change in formation energy is determined by the preference of the particle pair for the dislocation cores. This relation is also valid for balanced Janus particles, which are expelled from the cores at large size rather than bound to them.

In our analysis, we have considered one particle pair and one isolated dislocation pair. Our analysis does not address collective effects at finite particle concentration, including cooperative pinning, clustering, particle-mediated interactions, or the defect density in systems with many defects. Furthermore, we have imposed periodic boundary conditions, mimicking a bulk situation. In thin films, substrate interactions and free surface selectivity may change the picture. Our model represents particles as idealized spheres and resolves mesoscale thermodynamics rather than chain-level structure. The framework can be extended to anisotropic particles,\cite{zhangHarnessingAnisotropicNanoposts2014,tangPreciseControlPositioning2023, diazNematicOrderingAnisotropic2022} mixed particle types, finite particle concentrations, other defect types and curved or substrate selective films.\cite{Vu2018,qiaoStabilityElasticityUltrathin2024} The patch-ratio dependence also suggests a route to active defect control: particles with switchable surface affinities\cite{motornovStimuliResponsiveColloidalSystems2007} could alternate between homogeneous-like stabilization and Janus-like destabilization, enabling controlled defect nucleation, pinning, and release.\\


\begin{acknowledgments}
This research was funded by the Deutsche Forschungsgemeinschaft (DFG, Germany) -- Project numbers 465145163 (CRC 1552, project C02); 248882694; 464588647 (CRC 1551). The authors gratefully acknowledge the computing time granted on the MOGON NHR supercomputer at Johannes Gutenberg University Mainz (https://hpc.uni-mainz.de).
\end{acknowledgments}

\bibliography{Refs}
\clearpage
\onecolumngrid
\begin{center}
\textbf{\large Supporting Information}
\end{center}
\renewcommand{\thesection}{S\arabic{section}}
\setcounter{equation}{0}\renewcommand{\theequation}{S\arabic{equation}}
\setcounter{figure}{0}\renewcommand{\thefigure}{S\arabic{figure}}
\setcounter{section}{0}

\section{Numerical Implementation: Semi-Implicit Pseudospectral Method}
\label{app:numerics}

The order-parameter field $\psi(\mathbf{r},t)$ is evolved with a semi-implicit
pseudospectral scheme for the conserved time-dependent Ginzburg--Landau equation
of Cahn--Hilliard type. Spatial derivatives and the nonlocal chain-connectivity
term are evaluated in Fourier space using FFTW, while the local nonlinearities,
the particle coupling and the masked long-range term are computed in real space.
This combines numerical stability for the stiff linear terms with an efficient
treatment of the nonlocal interaction.

\subsection{Governing equation}
The dynamics of the order parameter are governed by
\begin{equation}
\frac{\partial \psi(\mathbf{r},t)}{\partial t}
= M \nabla^2 \mu(\mathbf{r},t),
\label{eq:si_chc}
\end{equation}
where $M$ is the mobility and $\mu=\delta \mathcal{F}/\delta \psi$ is the
chemical potential. The free-energy functional is Eq.~\eqref{eq:free_energy} of
the main text,
\begin{equation}
\begin{aligned}
\mathcal{F}[\psi,\{\mathbf{r}_i,\mathbf{u}_i\}]
&=\int d\mathbf{r}\left[H(\psi)+\frac{D}{2}(\nabla \psi)^2\right]\\
&\quad+ \frac{B}{2}\int d\mathbf{r}\int d\mathbf{r}'\,
G(\mathbf{r}-\mathbf{r}')\,
\widetilde{\psi}(\mathbf{r})\,\widetilde{\psi}(\mathbf{r}')
+ \mathcal{F}_{cp} + U_{cc},
\end{aligned}
\label{eq:si_free_energy}
\end{equation}
with the local free-energy density
\begin{equation}
H(\psi)=\frac{\tau_0}{2}\psi^2+\frac{v}{3}(1-2f)\psi^3+\frac{u}{4}\psi^4 ,
\label{eq:si_local_free_energy}
\end{equation}
and $\widetilde{\psi}(\mathbf{r})=\psi(\mathbf{r})\prod_i\Lambda_c(\mathbf{r}-\mathbf{r}_i)$
the masked field of the main text, with $\Lambda_c$ equal to zero inside the
hard core of any particle and unity elsewhere.

Taking the functional derivative, and noting that differentiating the double
integral produces two identical terms that cancel the factor $1/2$, gives
\begin{equation}
\mu(\mathbf{r},t)
= -D\nabla^2\psi
+\frac{\partial H}{\partial \psi}
+ B\,\Lambda_c(\mathbf{r})\!\int d\mathbf{r}'\,
G(\mathbf{r}-\mathbf{r}')\,\widetilde{\psi}(\mathbf{r}')
+ \mu_c(\mathbf{r},t),
\qquad
\mu_c \equiv \frac{\delta \mathcal{F}_{cp}}{\delta \psi}
= 2g_{cp}\sum_{i}\psi_c\!\left(|\mathbf{r}-\mathbf{r}_i|\right)
\left[\psi(\mathbf{r})-\psi_0(\mathbf{r},\mathbf{u}_i)\right].
\label{eq:si_chemical_potential}
\end{equation}
Here $\Lambda_c(\mathbf{r})$ abbreviates $\prod_i\Lambda_c(\mathbf{r}-\mathbf{r}_i)$.
The mask therefore appears twice: once inside the integral, so that polymer
displaced by a particle does not act as a source, and once outside it, so that
the same region does not respond to the potential generated by the surrounding
melt. Retaining only the inner factor would apply the long-range potential
inside the particle cores, since convolution with $G$ spreads the potential over
regions where the source vanishes.

\subsection{Torque from the Particle--Polymer Coupling}
\label{app:torque}

For a patchy particle the coupling free energy is
\begin{equation}
\mathcal{F}_{cp}
=g_{cp}\sum_{i=1}^{N_c}\int d\mathbf{r}\;
\psi_c\!\left(|\mathbf{r}-\mathbf{r}_i|\right)
\left[\psi(\mathbf{r})-\psi_0(\mathbf{r},\mathbf{u}_i)\right]^{2},
\end{equation}
with the preferred local composition
\begin{equation}
\psi_0(\mathbf{r},\mathbf{u}_i)
=\psi_b+(\psi_a-\psi_b)\,
H_s\!\left(\hat{\mathbf{r}}_i\cdot\mathbf{u}_i-\cos(\pi\alpha)\right),
\qquad
H_s(x)=\tfrac12\left[1+\tanh(k_s x)\right],
\end{equation}
and $\hat{\mathbf{r}}_i=(\mathbf{r}-\mathbf{r}_i)/|\mathbf{r}-\mathbf{r}_i|$.
Since the orientation is constrained to unit length, the physical torque is the
projection of the orientational derivative onto the tangent plane of
$\mathbf{u}_i$,
\begin{equation}
\mathbf{T}_i=-\,\mathbf{u}_i\times
\frac{\partial \mathcal{F}_{cp}}{\partial \mathbf{u}_i} .
\end{equation}
Differentiating gives
\begin{equation}
\frac{\partial \mathcal{F}_{cp}}{\partial \mathbf{u}_i}
=-2g_{cp}\int d\mathbf{r}\;
\psi_c\!\left(|\mathbf{r}-\mathbf{r}_i|\right)
\left[\psi(\mathbf{r})-\psi_0\right]
\frac{\partial \psi_0}{\partial \mathbf{u}_i},
\qquad
\frac{\partial \psi_0}{\partial \mathbf{u}_i}
=(\psi_a-\psi_b)\frac{k_s}{2}
\operatorname{sech}^{2}\!\left(k_s\left[\hat{\mathbf{r}}_i\cdot\mathbf{u}_i-\cos(\pi\alpha)\right]\right)\hat{\mathbf{r}}_i,
\end{equation}
so that
\begin{equation}
\mathbf{T}_i
=g_{cp}k_s(\psi_a-\psi_b)\int d\mathbf{r}\;
\psi_c\!\left(|\mathbf{r}-\mathbf{r}_i|\right)
\left[\psi(\mathbf{r})-\psi_0\right]
\operatorname{sech}^{2}\!\left(k_s\left[\hat{\mathbf{r}}_i\cdot\mathbf{u}_i-\cos(\pi\alpha)\right]\right)
\mathbf{u}_i\times\hat{\mathbf{r}}_i .
\end{equation}
The torque vanishes where the local composition matches the preferred value and
where the smooth step is saturated, and is largest near the patch boundary,
where the derivative of $H_s$ peaks. For homogeneous particles $\psi_0$ is
independent of $\mathbf{u}_i$ and the torque vanishes identically.

\subsection{Rodrigues' Rotation Formula}
\label{app:rodrigues}

The particle orientation is updated from the angular velocity using Rodrigues'
rotation formula. For an orientation $\mathbf{u}$ rotated by an angle $\theta$
about a unit axis $\mathbf{a}$,
\begin{equation}
\mathbf{u}' = \mathbf{u}\cos\theta + (\mathbf{a}\times\mathbf{u})\sin\theta
+ \mathbf{a}(\mathbf{a}\cdot\mathbf{u})(1-\cos\theta).
\end{equation}
In the overdamped rotational dynamics used here the angular velocity is
$\boldsymbol{\omega}=(\mathbf{T}\times\mathbf{u})/\gamma_{r}$, which is
perpendicular to $\mathbf{u}$. With
$\boldsymbol{\theta}=\boldsymbol{\omega}\Delta t_c$, $\theta=|\boldsymbol{\theta}|$
and $\mathbf{a}=\boldsymbol{\theta}/\theta$, it follows that
$\mathbf{a}\cdot\mathbf{u}=0$, the last term vanishes, and the update reduces to
\begin{equation}
\mathbf{u}' = \mathbf{u}\cos\theta + (\mathbf{a}\times\mathbf{u})\sin\theta .
\end{equation}
The orientation is renormalized after each update to remove accumulated
numerical drift. This finite-angle update is used directly, rather than the
small-angle approximation
$\mathbf{u}'\approx\mathbf{u}+\boldsymbol{\theta}\times\mathbf{u}$.

\subsection{Fourier-space formulation}

Expanding the order parameter in Fourier modes,
$\psi(\mathbf{r},t)=\sum_{\mathbf{k}}\hat\psi(\mathbf{k},t)e^{i\mathbf{k}\cdot\mathbf{r}}$,
the Laplacian becomes $\nabla^2\psi\rightarrow-k^2\hat\psi(\mathbf{k},t)$ with
$k^2=k_x^2+k_y^2+k_z^2$. The Green's function of $-\nabla^2$ satisfies
$\nabla^2G(\mathbf{r}-\mathbf{r}')=-\delta(\mathbf{r}-\mathbf{r}')$, so that
\begin{equation}
\hat{G}(\mathbf{k})=
\begin{cases}
1/k^{2}, & \mathbf{k}\neq 0,\\[2pt]
0, & \mathbf{k}=0 .
\end{cases}
\label{eq:si_green_fourier}
\end{equation}
Setting the $\mathbf{k}=0$ component to zero is the standard neutralizing
background of the Ohta--Kawasaki functional and is equivalent to replacing the
masked field by its deviation from the box average,
$\widetilde{\psi}\rightarrow\widetilde{\psi}-\overline{\widetilde{\psi}}$. This
step cannot be omitted here: although $\psi$ itself has zero mean, the masked
field $\widetilde{\psi}$ does not, because the mask removes material
preferentially from one block, and the $\mathbf{k}=0$ mode of $\hat G$ would
otherwise diverge.

Writing the explicit real-space contributions as
\begin{equation}
N(\mathbf{r},t)=v(1-2f)\psi^{2}+u\psi^{3}+\mu_{c}
+ B\,\Lambda_c(\mathbf{r})\,\Phi(\mathbf{r},t),
\qquad
\Phi \equiv \mathrm{FT}^{-1}\!\left[\hat{G}(\mathbf{k})\,
\widehat{\widetilde{\psi}}(\mathbf{k},t)\right],
\label{eq:si_nonlinear}
\end{equation}
where $\mathrm{FT}^{-1}$ denotes the inverse Fourier transform, the linear term
$\tau_0\psi$ is kept separate for implicit treatment. Note that the long-range
term enters $N$ rather than being applied as a multiplication in Fourier space:
the product $\Lambda_c(\mathbf{r})\Phi(\mathbf{r})$ is local in real space and
therefore a convolution in Fourier space, so $\Phi$ must be transformed back
before the mask is applied. The Fourier-space evolution equation is then
\begin{equation}
\frac{\partial \hat{\psi}(\mathbf{k},t)}{\partial t}
= -Mk^{2}\left[(Dk^{2}+\tau_0)\hat{\psi}(\mathbf{k},t)
+ \hat{N}(\mathbf{k},t)\right].
\label{eq:si_fourier_pde}
\end{equation}
A thermal noise term would enter on the right-hand side; its amplitude is set to
zero in all simulations reported here, so the field evolution is deterministic.

\subsection{Semi-implicit time integration}

The linear terms proportional to $(Dk^{2}+\tau_0)\hat\psi$ are treated
implicitly and the remaining contributions explicitly at time step $n$, giving
the first-order update
\begin{equation}
\hat{\psi}^{\,n+1}(\mathbf{k})
=\frac{\hat{\psi}^{\,n}(\mathbf{k})
-\Delta t\,Mk^{2}\,\hat{N}^{\,n}(\mathbf{k})}
{1+\Delta t\,Mk^{2}\left(Dk^{2}+\tau_0\right)} ,
\label{eq:si_update}
\end{equation}
where $\Delta t$ is the field time step. This treatment relaxes the stability
constraint associated with the biharmonic term and permits substantially larger
time steps than an explicit scheme, while retaining spectral accuracy for the
relaxation of lamellar interfaces and defect cores.

\subsection{Algorithmic procedure}

Each time step proceeds as follows.
\begin{enumerate}
    \item Update the particle fields $\psi_c$, $\psi_0$ and $\Lambda_c$ in real
          space from the current particle positions and orientations.
    \item Compute the local nonlinear terms and the coupling contribution
          $\mu_c$ in real space, and form the masked field $\widetilde{\psi}$.
    \item Transform $\psi$ and $\widetilde{\psi}$ to Fourier space.
    \item Multiply $\widehat{\widetilde{\psi}}$ by $\hat G(\mathbf{k})$,
          Eq.~\eqref{eq:si_green_fourier}, transform back to obtain
          $\Phi(\mathbf{r})$, apply the outer mask, and add
          $B\Lambda_c\Phi$ to the explicit term $N$,
          Eq.~\eqref{eq:si_nonlinear}.
    \item Transform $N$ to Fourier space and update $\hat\psi$ according to
          Eq.~\eqref{eq:si_update}.
    \item Transform the updated field back to real space.
    \item Advance particle positions and orientations with the overdamped translational and rotational Langevin equations of the main text.
\end{enumerate}

\section{Particle-Pair Separation}
\label{si:separation}

\begin{figure*}[t]
\centering
\includegraphics[width=0.9\linewidth]{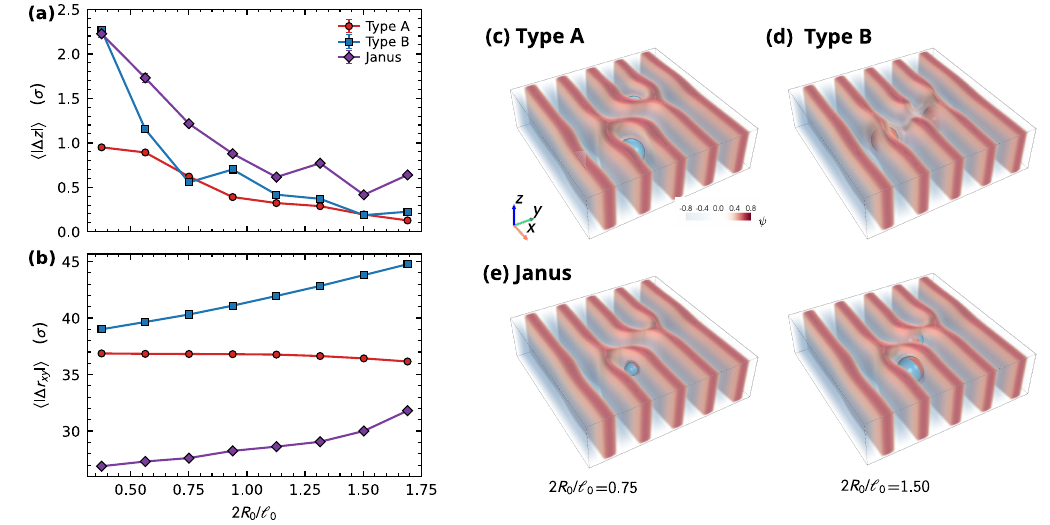}
\caption{Equilibrium particle-pair separation in the defect-containing state. (a)~Mean out-of-plane separation $\langle|\Delta z|\rangle$ and (b)~mean
in-plane separation $\langle|\Delta r_{xy}|\rangle$ of the two particles, versus reduced particle size $2R_0/\ell_0$, for Type~A (red circles), Type~B (blue squares) and balanced Janus ($\alpha=0.5$, purple diamonds) particles. Both quantities are averaged over the production stage and over the five commensurate lateral box sizes at $L_z=30\,\sigma$. (c--e)~Three-dimensional snapshots of the order-parameter field $\psi(\mathbf r)$ around the dislocation pair in the final production frame, for (c)~Type~A, (d)~Type~B and (e)~Janus particles; the left and right images in each panel show $2R_0/\ell_0=0.75$ and $1.50$ at $L_z=30\,\sigma$ and
$N=5$. Spheres mark the hard-core radius $R_0$ and are colored by particle type; the field is masked within the coupling radius $R_c=R_0+w$ around each particle so that the particles remain visible, and the mask is a rendering choice only. The $\psi$ color scale is shared with Fig.~\ref{fig:sketch}(b).}
\label{SI_fig:separation}
\end{figure*}
Figure~\ref{SI_fig:separation} shows the equilibrium particle-pair separations in the defect-containing state. Both particles are initially placed at the mid-plane, $z=L_z/2$; any nonzero out-of-plane separation is therefore generated during relaxation. For homogeneous particles, $\langle|\Delta z|\rangle$ decreases rapidly with particle size and drops below $0.5\,\sigma$ for the largest particles, since the coupling free energy carries no orientational degree of freedom and is symmetric in $z$ for these particles, so there is no systematic force driving them out of the mid-plane. The particles thus remain approximately coplanar. In contrast, the vertical separation of Janus particles is much larger for all particle sizes considered. In fact, inspection of $\langle (\Delta z)^2 \rangle$ as a function of time shows that Janus particles diffuse in the $z$ direction largely independent of each other, whereas homogeneous particles are bound. The in-plane separation $\langle|\Delta r_{xy}|\rangle$ varies with particle type and size (Fig.~\ref{SI_fig:separation}(b)). Together with the three-dimensional fields in Fig.~\ref{SI_fig:separation}(c--e), this demonstrates that particle type controls not only whether particles bind to the cores, but also their three-dimensional arrangement relative to the dislocation pair. Homogeneous particles localize near the cores in approximately coplanar configurations, whereas Janus particles move more independently from each other because their surface preferences couple to the local interfacial orientation.
\section{Finite-Size Scaling: Defect Formation Energy Extrapolation}
\label{SI_sec:fitting}
All formation and binding energies quoted in the main text are
thermodynamic-limit values obtained by extrapolating a finite-size sequence with Eq.~\eqref{eq:fss-general}. For each number of lamellar periods $N$ the lateral box size is fixed to the commensurate value $L^{*}_{l}(N)$ determined in Sec.~\ref{sec:bulk-results} and listed in Table~\ref{tab:si-boxes}. The same value is used for all four relaxed states at a given $N$, so that every free-energy difference is formed at identical box size, grid and particle
number.

\begin{figure*}[thb]
\centering
\includegraphics[width=\linewidth]{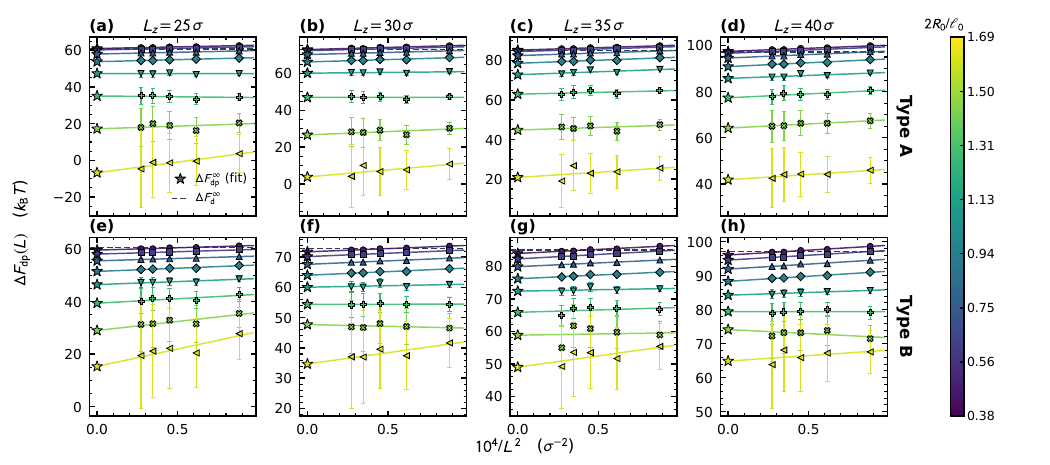}
\caption{Finite-size scaling of the dislocation-pair formation energy for
homogeneous particles. Panels (a--d) show Type~A and (e--h) Type~B particles, at
vertical periods $L_z=25$, $30$, $35$ and $40\,\sigma$ from left to right.
Symbols are the measured sequence $\Delta F_{dp}(L)$ against $1/L^{2}$ for each
particle size (color, see colorbar); solid lines are least-squares fits to
Eq.~\eqref{eq:fss-general} and stars mark the extrapolated values
$\Delta F_{dp}^{\infty}$ at $1/L^{2}=0$. The dashed line in each panel is the
undoped reference $\Delta F_{d}^{\infty}=cL_z$. Error bars are the standard
errors of the individual free-energy measurements.}
\label{fig:homo_fit}
\end{figure*}

\begin{figure*}[t]
\centering
\includegraphics[width=\linewidth]{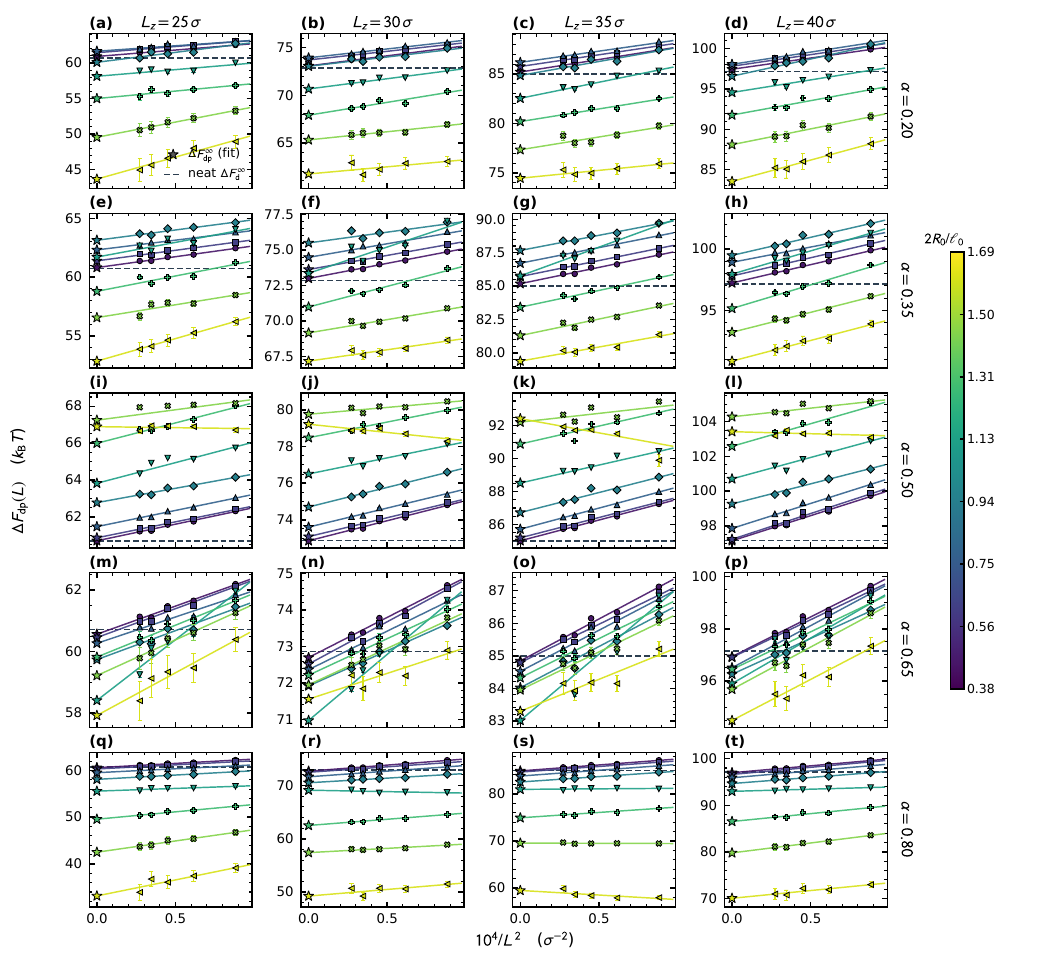}
\caption{Finite-size scaling of the dislocation-pair formation energy for patchy particles. Rows correspond to patch ratios $\alpha=0.20$, $0.35$, $0.50$, $0.65$ and $0.80$ from top to bottom, and columns to vertical periods $L_z=25$, $30$, $35$ and $40\,\sigma$ from left to right. Symbols, colors, fits, stars, dashed reference line and error bars are as in Fig.~\ref{fig:homo_fit}. The row $\alpha=0.50$ is the balanced Janus particle of Fig.~\ref{fig4:patchy-particles}.}
\label{fig:patchy_fit}
\end{figure*}

\begin{figure*}[t]
\centering
\includegraphics[width=\linewidth]{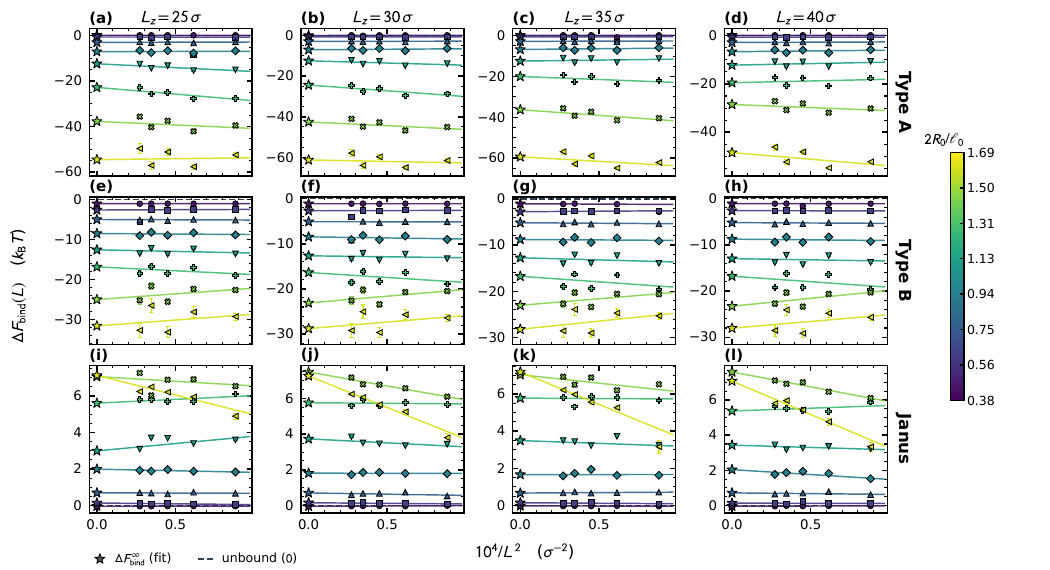}
\caption{Finite-size scaling of the particle--core binding free energy,
Eq.~\eqref{eq:dFbind}. Rows show Type~A (a--d), Type~B (e--h) and balanced Janus ($\alpha=0.5$, i--l) particles, and columns vertical periods $L_z=25$, $30$, $35$ and $40\,\sigma$ from left to right. Symbols are the measured sequence $\Delta F_{\mathrm{bind}}(L)$ against $1/L^{2}$ at the five commensurate box sizes of Table~\ref{tab:si-boxes}; color and symbol shape encode the reduced particle diameter $2R_0/\ell_0$ (colorbar, right). Solid lines are independent weighted least-squares fits to Eq.~\eqref{eq:fss-general} and stars mark the extrapolated values $\Delta F_{\mathrm{bind}}^{\infty}$ at $1/L^{2}=0$, which are plotted against particle size in Fig.~\ref{fig7:binding}(a). The dashed line marks the unbound reference $\Delta F_{\mathrm{bind}}=0$; Error bars are the standard errors of the individual free-energy measurements. Note the different vertical scale in each row.}
\label{fig:binding_fit}
\end{figure*}

\begin{table}[t]
\caption{Commensurate lateral box sizes $L^{*}_{l}(N)$ used for the finite-size extrapolation, with the corresponding lateral grid dimension and abscissa of Figs.~\ref{fig:homo_fit}--\ref{fig:binding_fit}.}
\label{tab:si-boxes}
\begin{ruledtabular}
\begin{tabular}{cccc}
$N$ & $L^{*}_{l}$ $(\sigma)$ & Lateral grid & $10^{4}/L^{2}$ $(\sigma^{-2})$ \\
\hline
5 & 106.5 & $213^2$ & 0.882 \\
6 & 127.5 & $255^2$ & 0.615 \\
7 & 149.0 & $298^2$ & 0.450 \\
8 & 170.0 & $340^2$ & 0.346 \\
9 & 191.5 & $383^2$ & 0.273 \\
\end{tabular}
\end{ruledtabular}
\end{table}
Free energies are averaged over the production stage, and the error bars in Figs.~\ref{fig:homo_fit}--\ref{fig:binding_fit} are the standard errors of these averages. Each sequence is fitted independently to Eq.~\eqref{eq:fss-general} by weighted least squares in $1/L^{2}$, separately for each surface chemistry, core radius $R_0$ and vertical period $L_z$, using the five points of Table~\ref{tab:si-boxes}. The uncertainty on $\Delta F^{\infty}$ is the standard error of the intercept from the fit covariance matrix, rescaled by the reduced chi-square, so that it reflects the scatter of the sequence about the linear form. Where a quantity is reported as a mean over $L_z$ in the main text, the error bar is instead the standard error across the four thicknesses.

\end{document}